\documentclass[journal]{IEEEtran}
\usepackage{amsmath,amssymb,amsfonts,bm}
\usepackage{array,booktabs}
\usepackage{algorithm}
\usepackage{algpseudocode}
\usepackage[caption=false,font=footnotesize]{subfig}
\usepackage{textcomp,stfloats,url,graphicx,cite}
\usepackage{xcolor,tikz,pgfplots}
\usepgfplotslibrary{groupplots}
\usepackage{helvet}
\usepackage[colorlinks=true,allcolors=black]{hyperref}
\pgfplotsset{compat=1.18}
\definecolor{originGray}{rgb}{.318,.318,.318}
\definecolor{originRed}{rgb}{.945,.251,.251}
\definecolor{originBlue}{rgb}{.102,.435,.875}
\definecolor{originGreen}{rgb}{.216,.678,.420}
\definecolor{originOrange}{HTML}{C88835}
\definecolor{originPurple}{HTML}{9563B5}
\pgfplotsset{originpaper/.style={
axis line style={black,line width=.8pt}, tick align=inside,tick style={black,line width=.65pt}, tick label style={font=\footnotesize\sffamily}, label style={font=\footnotesize\bfseries}, grid style={dashed,gray!65,line width=.3pt}, every axis plot/.append style={line width=1pt,mark size=2.3pt, mark options={solid,fill=white}}}}
\newcommand{\E}{\mathbb E}
\newcommand{\ind}{\mathbf 1}
\newcommand{\FAR}{FAR-MAPPO}
\begin{document}
\title{Feedback-Aware Residual Multi-Agent Reinforcement Learning for Timely Communication in Cognitive Radio Networks}
\author{Baolin Chong, Xiuqi Chen, Jiandong Guo, Fuhui Zhou,~\IEEEmembership{Senior Member,~IEEE}, and Qihui Wu,~\IEEEmembership{Fellow,~IEEE}

\IEEEcompsocitemizethanks{\IEEEcompsocthanksitem
Baolin Chong and Jiandong Guo are with the Key Laboratory of Unmanned Aerial Vehicle Technology of the Ministry of Industry and Information Technology, Nanjing University of Aeronautics and Astronautics, Nanjing 210016, China (e-mail: baolinchong@nuaa.edu.cn, bh4rbc@nuaa.edu.cn).

Xiuqi Chen is with the College of Automation Engineering, Nanjing University of Aeronautics and Astronautics, Nanjing 211106, China (e-mail:ricardocxq@nuaa.edu.cn)

Fuhui Zhou is with the College of Artificial Intelligence, Nanjing University of Aeronautics and Astronautics, Nanjing 210000, China (e-mail: zhoufuhui@ieee.org).

Qihui Wu is with the College of Electronic and Information Engineering, Nanjing University of Aeronautics and Astronautics, Nanjing 210000, China (e-mail: wuqihui2014@sina.com).
\protect\\}}
\maketitle
\begin{abstract}
Supporting time-sensitive services in cognitive radio networks requires distributed access strategies that translate shared spectrum opportunities into timely packet delivery. Achieving this objective is challenging because reception conditions are only partially observable to individual transmitters, while dynamic competition and access overhead constrain the service that can be completed within packet deadlines. We formulate a distributed access problem that captures the coupling among interference-dependent reception, queue evolution, and protocol durations. To address this problem, we propose feedback-aware residual multi-agent proximal policy optimization (FAR-MAPPO) under centralized training and decentralized execution. The method constructs channel-specific historical feedback representations that complement instantaneous sensing with evidence from completed access attempts. A residual policy network integrates this experience with current observations and packet urgency to guide local access decisions. Numerical results show improved timely delivery over the baseline methods, particularly under heavy traffic and bursty arrivals with mixed packet deadlines.
\end{abstract}
\begin{IEEEkeywords}
Cognitive radio networks, deadline-constrained communication, distributed channel access, historical feedback, multi-agent reinforcement learning.
\end{IEEEkeywords}

\section{Introduction}
\label{sec:intro}

\IEEEPARstart{T}{imely} communication is essential to time-sensitive wireless services~\cite{urllc}. Information with a limited useful lifetime must reach its destination before that lifetime expires. When multiple communication links share limited radio resources, achieving this objective requires terminals to adapt their access behavior to changing competition and reception conditions. Cognitive radio networks support such adaptation through environmental sensing, interaction feedback, and behavioral learning~\cite{haykin,survey}. Under distributed operation, however, translating locally available information into effective timely delivery remains challenging.

This challenge arises from the interaction between incomplete information and limited service time. Transmitter-side sensing cannot fully describe the interference experienced by the intended receiver, and an apparent access opportunity may disappear as competing activity changes. Meanwhile, obtaining information, acquiring access, and completing a transmission all consume the remaining lifetime of a packet. A channel that appears favorable at the decision instant may therefore fail to support timely delivery. This distinction becomes particularly consequential when traffic arrives in bursts and multiple links compete for few channels. Access decisions must account for both the reliability of a communication opportunity and the time required to exploit it.

Wireless access under dynamic competition has been studied extensively. Distributed learning and online allocation improve channel utilization under uncertain network conditions~\cite{naparstek,tan,convergence,orderoptimal}, while adaptive contention control addresses competition and coexistence~\cite{backoff,federated,coexistence}. Recent studies further support cooperation among heterogeneous users~\cite{harldsa,qpmix} and adaptation across network configurations~\cite{curriculum,multitask}. Historical sensing and reception feedback also help terminals infer hidden activity and exploit temporal dependencies~\cite{hidden,feedbackq,spiking}. These advances establish a foundation for access decisions based on local communication experience.
For time-sensitive services, existing work considers deadline-constrained access with switching overhead~\cite{xue} and distributed learning for timely industrial communication~\cite{industrial}. Traffic-aware resource allocation and access control further reduce packet delay or support differentiated service requirements~\cite{multicell,delayaccess,pdcf}. These studies highlight the importance of incorporating service needs into access design.

For autonomous transmitter--receiver links, a key unresolved issue is how to jointly account for receiver-dependent reception conditions and deadline consumption throughout access. Channel utilization and average delay alone do not determine whether an individual packet is delivered before expiry. Moreover, the access behavior of each link changes the interference experienced by other receivers, coupling their delivery opportunities. The scientific question is therefore how each link can use incomplete local evidence and communication experience to coordinate access when reception reliability and service completion time jointly determine timely delivery.

Multi-agent reinforcement learning (MARL) offers a promising approach to this problem. Each link can act as an agent that selects actions from local observations and learns how its behavior affects subsequent communication opportunities and service outcomes. Centralized training with decentralized execution supports cooperative learning from joint information while retaining local decision-making after deployment~\cite{albrecht}. Multi-agent proximal policy optimization (MAPPO) provides a suitable actor--critic framework~\cite{mappo}. Directly applying a general policy network, however, still requires extracting channel-specific experience from delayed, sparse, and unevenly sampled feedback and relating that experience to current traffic conditions. A recurrent network can exploit observation sequences, and explicitly organizing feedback for the access task provides a more direct representation of the evidence needed for timely decisions.

Two related design challenges follow. First, communication experience must remain informative under changing competition. Long-term outcomes describe relatively persistent access behavior, whereas recent outcomes help expose changes in competing activity. The amount and freshness of the available evidence also affect its relevance to the next decision. Second, that experience must be used according to current service needs. A channel with favorable historical outcomes may be unsuitable for a packet with little remaining time because switching and access incur additional delays. Conversely, a channel with poor recent outcomes may become usable as competition changes. The policy must therefore combine historical feedback, current sensing, and packet urgency to make state-dependent access decisions.

To address these challenges, we propose feedback-aware residual multi-agent proximal policy optimization (FAR-MAPPO). The method constructs channel-specific feedback representations that describe experienced access conditions through outcomes at multiple time scales, evidence freshness, and operation duration. These representations are combined with the current local state and supplied to a residual policy network~\cite{resnet}. A direct feature path retains encoded information, while learnable residual transformations represent interactions between communication experience and the current service state. A centralized critic supports cooperative training, and each deployed actor makes decisions using only its own observations and feedback memory. Together, these components provide a task-specific policy representation for distributed timely access.

The main contributions are summarized as follows.
\begin{enumerate}
    \item We formulate distributed timely access in a cognitive radio network with multiple transmitter--receiver links and external competition. The model jointly captures the mismatch between local sensing and receiver conditions, dynamic interference, and deadline consumption during access operations. The objective measures the fraction of offered packets whose delivery is confirmed before expiry.

    \item We develop FAR-MAPPO by integrating a channel-specific feedback representation with a residual policy network in the MAPPO framework. Multi-time-scale outcomes, evidence freshness, and access duration organize local communication experience, while the residual actor combines this experience with current service conditions for decentralized access decisions.

    \item We establish a packet-level evaluation that examines controlled-link population, external competition, channel availability, and offered traffic separately. Traffic-burstiness and deadline-configuration comparisons, component ablations, and delivery analysis across deadline classes assess the operating conditions and component effects relevant to the proposed design.
\end{enumerate}

The remainder of this paper is organized as follows. Section~\ref{sec:system} presents the system model and problem formulation. Section~\ref{sec:method} develops FAR-MAPPO. Section~\ref{sec:experiments} presents the performance evaluation, and Section~\ref{sec:conclusion} concludes the paper.

\noindent\textit{Notation:} Bold symbols denote vectors, and $\mathbb{R}^{d}$ denotes the $d$-dimensional real vector space. For a set $\mathcal S$, $|\mathcal S|$ denotes its cardinality. The operators $\mathbb E[\cdot]$ and $\Pr\{\cdot\}$ denote expectation and probability, respectively. The indicator $\mathbf{1}\{\cdot\}$ equals one when its argument is true and zero otherwise. The notation $[\mathbf x;\mathbf y]$ denotes vector concatenation, and $x_{0:t}$ denotes the sequence from time index $0$ through $t$. The operator $\operatorname{clip}(x,a,b)$ restricts $x$ to the interval $[a,b]$.
\section{System Model and Problem Formulation}
\label{sec:system}

\begin{figure*}[t]
\centering
\includegraphics[width=\textwidth]{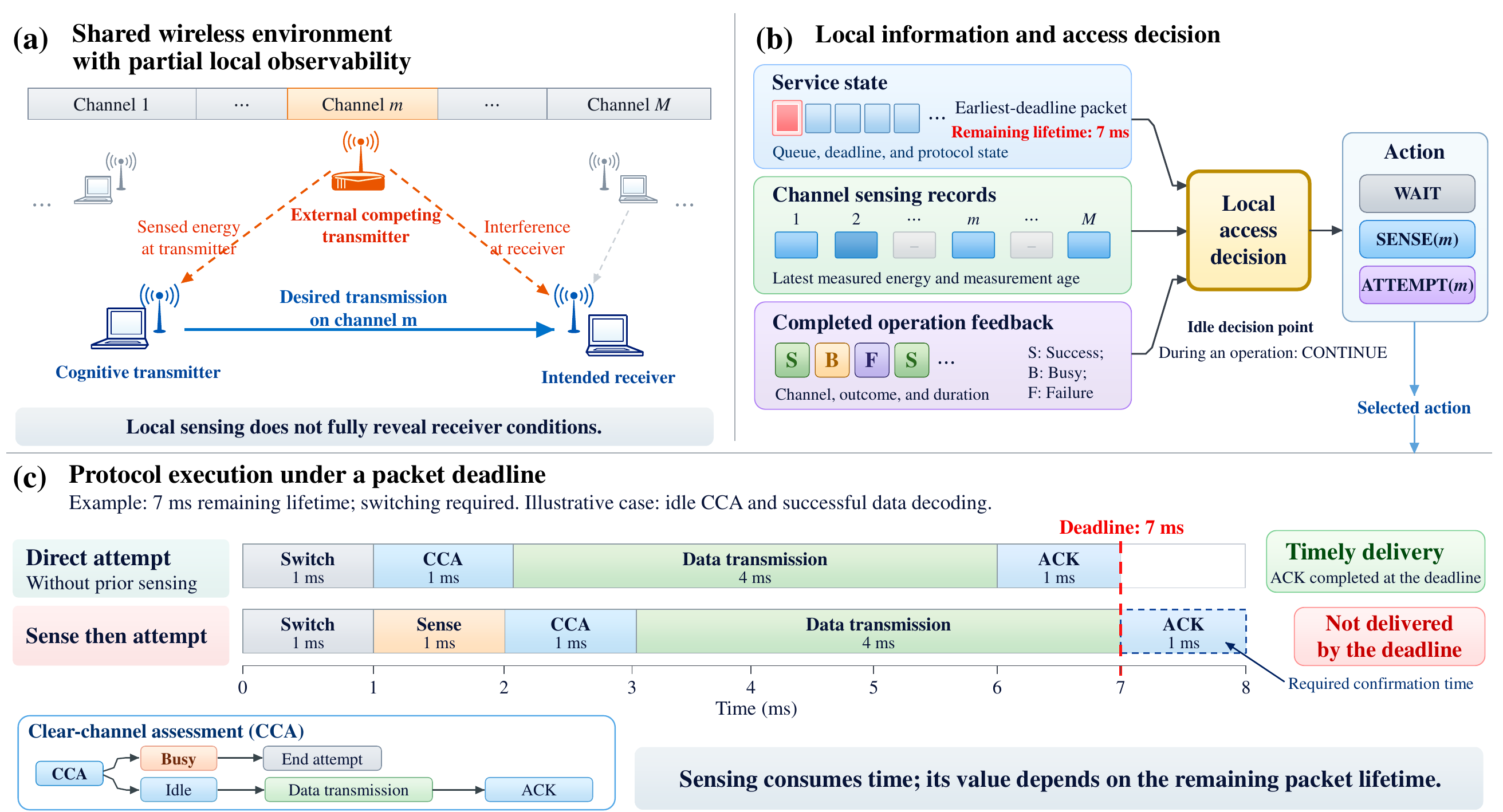}
\caption{System model: (a) shared wireless environment with partial local observability; (b) local information and access decisions; (c) protocol execution under a packet deadline.}
\label{fig:system}
\end{figure*}

We consider distributed timely communication among cognitive transmitter--receiver links sharing orthogonal channels with external competing links. As illustrated in Fig.~\ref{fig:system}, each controlled transmitter selects access actions from local service, sensing, and feedback information. The actions progress through protocol stages, alter transmission activity, and generate observations for subsequent decisions. The model below connects this local access loop to receiver conditions and packet deadlines.

\subsection{Network and Wireless Environment}

Consider $N$ controlled links and $K$ external competing links sharing $M$ orthogonal channels. Their index sets are $\mathcal N=\{1,\ldots,N\}$, $\mathcal K=\{N+1,\ldots,N+K\}$, and $\mathcal M=\{1,\ldots,M\}$, respectively. Each link consists of a transmitter and its intended receiver. A controlled link uses one channel at a time, and its receiver follows the channel selected by its transmitter. Time is divided into physical slots of duration $\Delta$, indexed by $t$. Unless otherwise stated, protocol durations and packet deadlines below are expressed in slots, and multiplication by $\Delta$ gives physical time. Link locations remain fixed within a realization, while fading and transmission activity evolve over time.

\paragraph{{Propagation and reception}}
Using a large-scale attenuation and block-fading model~\cite{goldsmith}, the channel power gain from transmitter $j$ to endpoint $x$ is
\begin{equation}
\label{eq:gain}
\begin{split}
g_{jx,m}(t)
={}&g_0
\left(
\frac{\max\{d_{jx},d_0\}}{d_0}
\right)^{-\alpha} 10^{\frac{\xi_{jx,m}}{10}}F_{jx,m}(t),
\end{split}
\end{equation}
where $d_{jx}$ is their distance, $d_0$ is the reference distance, $g_0$ is the reference power gain, and $\alpha$ is the path-loss exponent. The shadowing term $\xi_{jx,m}$ is Gaussian in decibels and fixed within a realization. The fading power $F_{jx,m}(t)$ is exponentially distributed with unit mean and remains constant over a coherence block. Shadowing and fading are sampled for the directed propagation paths and channels.

Let $\mathcal T_m(t)$ contain controlled and external transmitters in their data phase on channel $m$. With $g_{ji,m}(t)$ denoting the gain from transmitter $j$ to receiver $i$, the signal-to-interference-plus-noise ratio (SINR) of active controlled link $i$ is
\begin{equation}
\label{eq:sinr}
\Gamma_{i,m}(t)
=
\frac{P_i g_{ii,m}(t)}
{\sigma^2+
\sum_{j\in\mathcal T_m(t)\setminus\{i\}}
P_jg_{ji,m}(t)},
\end{equation}
where $P_j$ is the transmit power of link $j$ and $\sigma^2$ is the receiver noise power. A data transmission is decoded successfully if $\Gamma_{i,m}(t)\geq\Gamma_{\mathrm{th}}$ in every slot of its data phase, where $\Gamma_{\mathrm{th}}$ is the decoding SINR threshold in linear scale.

\paragraph{{Local sensing}}
Sensing measures aggregate power at the transmitter. Let $\widetilde g_{ji,m}(t)$ be the gain from transmitter $j$ to transmitter $i$. A measurement on channel $m$ is
\begin{equation}
\label{eq:energy}
\begin{split}
e_{i,m}(t)
={}&10\log_{10}\!
\left(
{
\sigma^2+
\sum\nolimits_{j\in\mathcal T_m(t)\setminus\{i\}}P_j\widetilde g_{ji,m}(t)}
\right)
+\varepsilon_{i,t},
\end{split}
\end{equation}
where $\varepsilon_{i,t}$ is zero-mean Gaussian measurement error in decibels. Energy is observed through sensing or clear-channel assessment (CCA) on the selected channel. Measurements of other channels retain their last values and associated ages. 

\paragraph{{External competition}}
Each external link is assigned a channel at initialization and follows an ON/OFF activity process with geometric state durations. During an ON period, it uses carrier sensing and random backoff to initiate data transmissions, following the carrier-sense/backoff principle analyzed in~\cite{bianchi}. A busy channel freezes its countdown, whereas an idle channel permits the countdown to advance. Its backoff window and transmission duration are fixed for that link within a realization. An ongoing data transmission completes even if the activity process becomes OFF. External transmissions enter both the sensed energy and receiver interference above, while controlled transmissions can influence external carrier sensing. External links supply autonomous competing traffic.

\subsection{Traffic and Deadline Model}

Let $A_i(t)\in\{0,1\}$ indicate an exogenous packet arrival at controlled link $i$ in slot $t$, with mean rate $\lambda_i$ packets per unit physical time. Under non-bursty traffic, arrivals are independent Bernoulli variables with probability $\lambda_i\Delta$. Under bursty traffic, a Markov state $Z_i(t)\in\{0,1\}$ modulates arrivals:
\begin{equation}
\label{eq:arrivals}
\Pr\{A_i(t)=1\mid Z_i(t)\}
=
\begin{cases}
\lambda_i\Delta/\varrho_i, & Z_i(t)=1,\\
0, & Z_i(t)=0,
\end{cases}
\end{equation}
where $\varrho_i=T_{\mathrm{on},i}/(T_{\mathrm{on},i}+T_{\mathrm{off},i})$ is the stationary ON probability and $T_{\mathrm{on},i}$ and $T_{\mathrm{off},i}$ are mean state durations in slots. The ON/OFF transition probabilities are $1/T_{\mathrm{on},i}$ and $1/T_{\mathrm{off},i}$, respectively. We initialize this process in stationarity and require $\lambda_i\Delta/\varrho_i\leq1$. Consequently, both traffic models have the same mean arrival rate, and their difference is temporal concentration of demand.

Packet $p$ arrives at slot $b_{i,p}$ with relative deadline $D_{i,p}$, independently drawn from a prescribed deadline distribution. Its absolute deadline and remaining slack are
\begin{equation}
\label{eq:deadline}
d_{i,p}=b_{i,p}+D_{i,p},
\qquad
s_{i,p}(t)=d_{i,p}-t.
\end{equation}
Each link has a queue $\mathcal Q_i(t)$ of capacity $B$. A new access attempt selects the queued packet with the earliest absolute deadline, with arrival order breaking ties, and retains that packet throughout the operation. New arrivals do not preempt an ongoing attempt. A packet remains counted in the queue while in service until it is delivered or expires. An unsuccessful attempt leaves an unexpired packet available for another attempt, and arrivals to a full queue are discarded.

Let $Y_i(t)$ be the number of timely deliveries confirmed at the end of slot $t$, and let $X_i(t)$ count the other queued packets that expire at that boundary. Confirmations are processed before expiry, so a successful confirmation exactly at the deadline is credited. Writing $q_i(t)=|\mathcal Q_i(t)|$, the queue-size evolution is
\begin{equation}
\label{eq:queue}
q_i(t+1)
=
\min\{
B,\, q_i(t)-Y_i(t)-X_i(t)+A_i(t+1)
\}.
\end{equation}
The corresponding excess arrivals are overflow losses. Timely delivery requires successful data reception and completion of the acknowledgment (ACK) phase by the packet deadline. The ACK phase provides reliable confirmation with a fixed duration and contributes no interference.

\subsection{Distributed Access and Local Observations}

\paragraph{{Local actions and protocol progression}}
Figs.~\ref{fig:system}(b) and~\ref{fig:system}(c) connect local decisions to their protocol execution. Let $a_{i,t}$ denote the action selected by controlled link $i$ at slot $t$. When idle, the link selects its action from the set
\begin{equation}
\label{eq:actions}
\begin{split}
\mathcal A_i
={}&\{\mathrm{WAIT}\}\\
&\cup
\{\mathrm{SENSE}(m),\mathrm{ATTEMPT}(m):
m\in\mathcal M\},
\end{split}
\end{equation}
giving $1+2M$ action indices. WAIT advances one slot. SENSE switches to the selected channel when necessary and then obtains an energy measurement. ATTEMPT selects a queued packet, switches when necessary, and performs CCA. A busy CCA ends the operation without a data transmission. If CCA is idle and the selected packet is still present and unexpired, the link enters its data phase and then the feedback phase. The CCA measurement also updates the local sensing record.

Let $m_i(t)$ be the current channel of link $i$. The stage durations $\tau_{\mathrm{sw}}$, $\tau_{\mathrm{sense}}$, $\tau_{\mathrm{CCA}}$, $\tau_{\mathrm{data}}$, and $\tau_{\mathrm{ACK}}$ denote the numbers of slots required for switching, sensing, CCA, data transmission, and ACK confirmation, respectively. The total sensing duration $\tau^{\mathrm{sense}}_{i,m}(t)$ and the duration $\tau^{\mathrm{att}}_{i,m}(t)$ of an uninterrupted attempt on channel $m$ that reaches confirmation are
\begin{align}
\tau^{\mathrm{sense}}_{i,m}(t)
&=
\tau_{\mathrm{sw}}\ind\{m\neq m_i(t)\}
+\tau_{\mathrm{sense}},
\label{eq:sense_duration}\\
\tau^{\mathrm{att}}_{i,m}(t)
&=
\tau_{\mathrm{sw}}\ind\{m\neq m_i(t)\}
+\tau_{\mathrm{CCA}}
+
\tau_{\mathrm{data}}
+\tau_{\mathrm{ACK}}.
\label{eq:duration}
\end{align}
Protocol stages proceed to completion once initiated, while packet lifetimes decrease on the same physical clock. Busy CCA or packet expiry before data initiation ends an attempt early. Entry into the data phase requires an unexpired packet, and evaluating whether its remaining lifetime can accommodate data and confirmation is part of the access decision.

While idle, WAIT and all sensing actions are legal, and attempt actions additionally require a nonempty queue. During an ongoing operation, CONTINUE advances the current protocol stage and is the sole legal action, using the same index as WAIT. All links act from pre-transition observations, after which interference is evaluated and ongoing stages advance on the shared physical clock. This accommodates overlapping operations and different decision opportunities across links.

\paragraph{{Information available to a terminal}}
The local observation comprises three groups:
\begin{equation}
\label{eq:observation}
o_{i,t}
=
\bigl(
o^{\mathrm{service}}_{i,t}, o^{\mathrm{sensing}}_{i,t}, o^{\mathrm{feedback}}_{i,t}
\bigr).
\end{equation}
The service group $o^{\mathrm{service}}_{i,t}$ includes link identity, current channel, protocol phase, queue length, earliest-deadline packet slack and attempt count, and the arrival indicator. The sensing group $o^{\mathrm{sensing}}_{i,t}$ records the latest measured energy, validity, age, and busy indication for each channel. The feedback group $o^{\mathrm{feedback}}_{i,t}$ contains the most recent initiated action, completed-operation outcome, and elapsed operation time. A packet selected by an ongoing attempt may differ from the earliest-deadline packet after a new arrival and the policy does not interrupt that attempt.

The actual receiver SINR, current external activity and backoff states, and the local observations of other links are unavailable to the deployed actor. Operation outcomes include successful timely delivery, unsuccessful delivery, busy assessment, and expiry-related termination. These outcomes summarize experienced access conditions across contention, reception, and packet expiry. Each completed operation generates one feedback event associated with its selected channel and duration. Section~\ref{sec:method} describes how the policy organizes this local experience.

\subsection{Problem Formulation}

Let $\mathcal C$ be the set of link--packet pairs $(i,p)$ generated during a fixed measurement window, including packets discarded on arrival. Let $\pi=\{\pi_i\}_{i\in\mathcal N}$ denote the collection of decentralized policies, where $\pi_i$ is the local policy of link $i$. Define $c_{i,p}$ as the successful confirmation time and set $c_{i,p}=+\infty$ if the packet is never successfully confirmed. The timely-delivery objective is
\begin{equation}
\label{eq:tdr}
\eta(\pi)
=
\E_\pi\!
\left[
\frac{
\sum_{(i,p)\in\mathcal C}
\ind\{c_{i,p}\leq d_{i,p}\}}
{|\mathcal C|}
\,\middle|\,
|\mathcal C|>0
\right],
\end{equation}
where $\eta(\pi)$ is the expected timely-delivery ratio, and $\E_\pi$ averages over the system randomness and actions generated under policy $\pi$. The denominator includes all offered packets, and the numerator counts their timely confirmations. Packets are followed through delivery or {expiry}, and traffic and competition continue throughout this observation period.

Let $\mathcal I_{i,t}=(o_{i,0:t},a_{i,0:t-1})$ be the causal local history of link $i$. Define the joint action vector $\mathbf a_t=[a_{1,t};\ldots;a_{N,t}]$ and the collection of local histories $\boldsymbol{\mathcal I}_t=(\mathcal I_{1,t},\ldots,\mathcal I_{N,t})$. 
For an action $a\in\mathcal A_i$, let $\mu_{i,t}(a)\in\{0,1\}$ denote the legal-action mask.
The quantity $\pi_i(a_{i,t}\mid\mathcal I_{i,t})$ is the conditional probability of the local action. We seek decentralized policies satisfying
\begin{align}
\label{eq:problem}
\max_{\{\pi_i\}_{i\in\mathcal N}}\quad
&\eta(\pi),\\
\text{s.t.}\quad
&\pi(\mathbf a_t\mid\boldsymbol{\mathcal I}_t)
=
\prod_{i\in\mathcal N}
\pi_i(a_{i,t}\mid\mathcal I_{i,t}),
\nonumber\\
&\pi_i(a\mid\mathcal I_{i,t})=0
\quad\text{if }\mu_{i,t}(a)=0,
\nonumber\\
&\sum_{a\in\mathcal A_i}
\pi_i(a\mid\mathcal I_{i,t})=1,
\nonumber\\
&\pi_i(a\mid\mathcal I_{i,t})\geq0,
\nonumber
\end{align}
for every link and decision time, under the physical, traffic, and protocol transitions above. The policies choose when to wait, obtain information, or attempt service and which channel to use. Packet service order, transmit power, and external access rules are fixed components of the system model.

Three coupled properties make this problem difficult. First, independent actions affect one another through receiver interference and carrier sensing, so a locally favorable choice may reduce the service opportunities of other links. Second, the observations provide incomplete and delayed evidence of reception conditions. Third, an action consumes time and changes the queue and deadline conditions under which later actions are taken. An instantaneous energy score captures only part of these consequences, while a fixed allocation cannot respond to changes in competition and demand. Direct model-based optimization would require a sufficiently informative state and a tractable transition model for the interacting links.

MARL can learn these temporally extended consequences from interaction while maintaining local execution. Centralized training supplies a critic with the joint observations of the controlled links, and each deployed actor conditions only on its own information. Effective learning still requires a representation that relates experienced access outcomes to present service urgency. 

\section{Feedback-Aware Residual Multi-Agent Reinforcement Learning}
\label{sec:method}

\begin{figure*}[t]
\centering
\includegraphics[width=\textwidth]{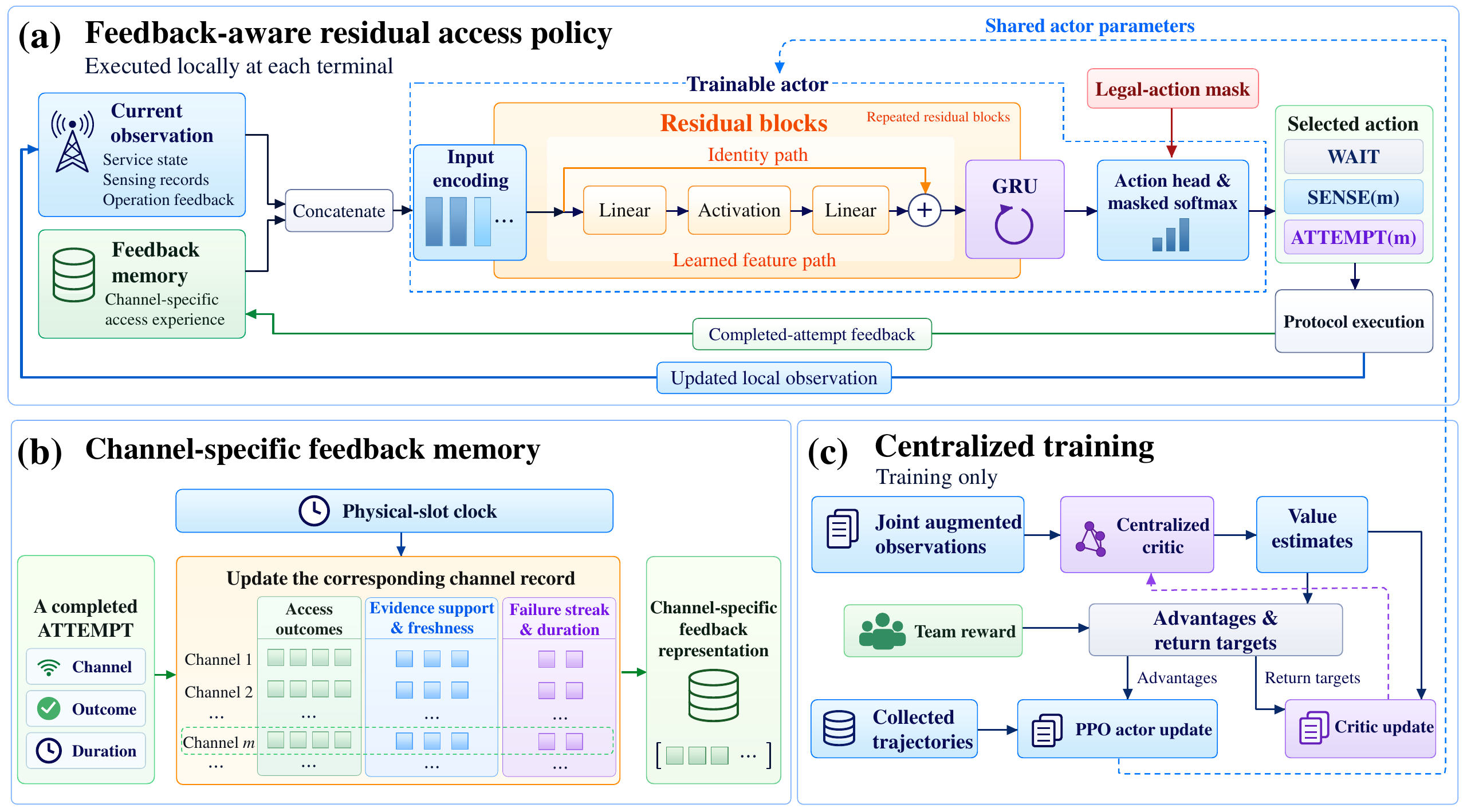}
\caption{Architecture of \FAR: (a) local feedback-aware residual access policy; (b) channel-specific feedback memory; (c) centralized training.}
\label{fig:method}
\end{figure*}

This section develops \FAR\ to solve the distributed timely-access problem in~\eqref{eq:problem}. The method combines a channel-specific representation of local feedback with a residual actor that interprets this experience under current service conditions. MAPPO provides the cooperative training framework~\cite{mappo}. As illustrated in Fig.~\ref{fig:method}, each terminal maintains a local feedback memory and executes a recurrent residual policy, while a centralized critic supports training through a shared timely-delivery objective.

\subsection{Framework Overview}

The communication system defines a cooperative decentralized partially observable Markov decision process~\cite{albrecht}. Its latent state comprises propagation conditions, external activity, queues, and protocol stages. Each agent acts on its local observation. The shared reward counts timely confirmations in physical slot $t$:
\begin{equation}
\label{eq:reward}
r_t
=
\sum_{i\in\mathcal N}Y_i(t).
\end{equation}
Using the discount factor $\gamma\in[0,1)$, the training objective accumulates these rewards over time:
\begin{equation}
\label{eq:discounted_return}
J_\gamma(\pi)
=
\E_\pi
\left[
\sum_{t\geq0}\gamma^t r_t
\right].
\end{equation}
The discounted return supplies the learning objective, while~\eqref{eq:tdr} measures finite-cohort timely delivery. Confirmation rewards assign value to earlier waiting, sensing, and access decisions through their subsequent delivery outcomes.

At each physical slot, terminal $i$ receives the local observation $o_{i,t}$ defined in~\eqref{eq:observation}. The local memory in Fig.~\ref{fig:method}(a) incorporates newly completed attempts and produces a feedback vector $\mathbf h_{i,t}$. The augmented policy input is
$\mathbf x_{i,t}
=
[\mathbf o_{i,t};\mathbf h_{i,t}],$
where $\mathbf o_{i,t}$ is the numerical vector encoding of the local observation $o_{i,t}$, and brackets denote concatenation. The actor encodes this input, applies residual feature transformations, and updates its recurrent state before selecting a legal action. Each terminal maintains its own memory and recurrent state, while the actor parameters, collectively denoted by $\theta$, are shared across controlled links. Link identity in the observation allows a shared policy to distinguish terminals.

{The feedback representation addresses an information problem: local sensing provides an incomplete account of receiver conditions, whereas completed attempts reveal whether access actually produced timely service. The residual actor addresses a representation problem: the relevance of that experience depends on current channel measurements, queue conditions, and deadline slack. A centralized critic estimates the future team delivery return during training. During execution, each terminal uses only its local input, memory, and actor, and no joint-observation exchange is required.}

\subsection{Historical Feedback Representation}

{Fig.~\ref{fig:method}(b)} illustrates how completed-attempt feedback updates the corresponding channel record to form the feedback representation.

\paragraph{{Event association and causal updates}}
A terminal records the selected channel and start time whenever it initiates ATTEMPT. When the operation completes, its outcome and elapsed duration are assigned to that channel, and the pending record is cleared. Attempt statistics are updated exactly once for each completed ATTEMPT event.

Let $u_{i,m,t+1}=1$ when an attempt on channel $m$ completes during the transition from $t$ to $t+1$, and zero otherwise. For a completed attempt, $y_{i,m,t+1}=1$ denotes timely confirmed delivery and $v_{i,m,t+1}=1$ denotes busy CCA, and each indicator is zero otherwise. Both outcome indicators are set to zero when no attempt completes. An unsuccessful outcome includes transmission failure, busy assessment, or expiry-related termination. The statistics below therefore characterize the empirical success of access attempts in achieving timely delivery, jointly reflecting contention, reception conditions, and service duration.

\paragraph{{Cumulative and recent outcomes}}
For each link--channel pair, initialize smoothed success and failure counts as $G_{i,m,0}=B_{i,m,0}=2$ and update
\begin{equation}
    \begin{aligned}
        \label{eq:history_counts}
G_{i,m,t+1}
&=
G_{i,m,t}
+
u_{i,m,t+1}y_{i,m,t+1},\\
B_{i,m,t+1}
&=
B_{i,m,t}
+
u_{i,m,t+1}(1-y_{i,m,t+1}).
    \end{aligned}
\end{equation}
The first update adds a successful completion, while the second adds an unsuccessful completion. Both counts remain unchanged when $u_{i,m,t+1}=0$. Their ratio defines the smoothed cumulative success statistic:
\begin{equation}
\label{eq:history_cumulative_rate}
\widehat p_{i,m,t}
=
\frac{G_{i,m,t}}
{G_{i,m,t}+B_{i,m,t}}.
\end{equation}
The actual number of observed attempts is
$C_{i,m,t}=G_{i,m,t}+B_{i,m,t}-4$,
where subtracting four removes the initial smoothing counts.

To expose changes on different time scales, maintain discounted success and attempt counts {$S^{(r)}_{i,m,t}$} and $C^{(r)}_{i,m,t}$, where $r\in\{f,s\}$ denotes fast and slow memory. Let $T_f<T_s$ be decay time constants expressed in the same physical-time units as $\Delta$, and set $\alpha_r=\exp(-\Delta/T_r)$. The two counts evolve as
\begin{equation}
    \begin{aligned}
        \label{eq:history_decay}
S^{(r)}_{i,m,t+1}
&=
\alpha_r S^{(r)}_{i,m,t}
+
u_{i,m,t+1}y_{i,m,t+1},\\
C^{(r)}_{i,m,t+1}
&=
\alpha_r C^{(r)}_{i,m,t}
+
u_{i,m,t+1}.
    \end{aligned}
\end{equation}
The first count tracks recent successes, and the second tracks all recent attempts. Both start at zero and decay every physical slot, including slots without completed attempts. The corresponding success statistic is
\begin{equation}
\label{eq:history_recent_rate}
\widehat p^{(r)}_{i,m,t}
=
\frac{2+S^{(r)}_{i,m,t}}
{4+C^{(r)}_{i,m,t}}.
\end{equation}
The smoothing constants give an initial value of $1/2$. As a channel remains unsampled, its discounted evidence loses weight and the statistic gradually returns toward this initial value.

To record recent contention separately, a busy count follows
\begin{equation}
\label{eq:history_busy}
U_{i,m,t+1}
=
\alpha_f U_{i,m,t}
+
u_{i,m,t+1}v_{i,m,t+1}.
\end{equation}
Here, $U_{i,m,0}=0$, and a completed attempt adds one only when CCA reports a busy channel. Normalizing by the discounted attempt count yields the busy-feedback statistic
\begin{equation}
\label{eq:history_busy_rate}
\widehat b_{i,m,t}
=
\frac{U_{i,m,t}}
{1+C^{(f)}_{i,m,t}}.
\end{equation}

\paragraph{{Evidence freshness and service duration}}
The memory also stores the last completion time $t^{\rm last}_{i,m}$, the consecutive unsuccessful-attempt count $L_{i,m,t}$, and a smoothed operation duration $\bar\tau_{i,m,t}$. At a completed successful attempt, the streak is reset to zero, and at any other completed attempt, it increases by one up to a cap $L_{\max}$.

If an operation started at $t^{\rm start}$ and completed at $t+1$, its observed duration is $\tau^{\rm obs}=t+1-t^{\rm start}$. The smoothed duration is updated as
\begin{equation}
\label{eq:history_duration}
\bar\tau_{i,m,t+1}
=
(1-\omega)\bar\tau_{i,m,t}
+
\omega\tau^{\rm obs},
\end{equation}
with $\omega=0.2$. Duration and streak remain unchanged between completion events. The initial duration is the nominal CCA--data--confirmation time on the current channel, and the initial streak is zero.

Feedback age is $A^{\rm fb}_{i,m,t}=t-t^{\rm last}_{i,m}$ after at least one completion. An unobserved channel is assigned the maximum normalized age. This quantity allows the policy to distinguish recently observed access conditions from older experience.

\paragraph{{Compact feature construction}}
To place counts, ages, and durations on a common scale, normalize a nonnegative scalar $z$ using a positive cap $z_{\max}$:
\begin{equation}
\label{eq:history_normalization}
\widetilde z
=
\operatorname{cap}(z;z_{\max})
=
\frac{\min\{z,z_{\max}\}}{z_{\max}}.
\end{equation}
Apply this rule to $C$, $C^{(f)}$, $A^{\rm fb}$, $L$, and $\bar\tau$, using the respective caps $C_{\max}$, $C^{(f)}_{\max}$, $A_{\max}$, $L_{\max}$, and $\tau_{\max}$. Since $L$ is already capped, its normalized value is $L/L_{\max}$.

For compact notation, suppress the common indices $(i,m,t)$ on the right-hand side below. The channel representation is
\begin{equation}
\label{eq:history}
\begin{aligned}
\mathbf f_{i,m,t}
=
\bigl[
&\widehat p;\ \widehat p^{(f)};\
\widehat p^{(s)};\ \widehat b;\widetilde C;\ \widetilde C^{(f)};\
\widetilde A^{\rm fb};\ \widetilde L;\
\widetilde\tau
\bigr]
\end{aligned}
\end{equation}
The first four entries summarize access outcomes, the next three describe sample support and freshness, and the last two record consecutive failures and experienced time cost.

Concatenating the channel vectors produces the feedback input of the actor:
\begin{equation}
\label{eq:history_concat}
\mathbf h_{i,t}
=
[\mathbf f_{i,1,t};\ldots;\mathbf f_{i,M,t}]
\in\mathbb R^{9M}.
\end{equation}
The memory resets with a new environment realization and is constructed entirely from the causal interaction record of the terminal.

These features serve complementary purposes. Cumulative outcomes summarize longer-term access experience, and disagreement with recent outcomes makes a change in observed performance visible to the actor. Counts and age distinguish repeatedly supported observations from sparse or stale evidence. Busy feedback and failure streaks describe recent access difficulties. Duration records the time actually consumed by attempts, including attempts that terminate early. Together with the current protocol state and deadline slack, it provides evidence of the time cost of accessing a channel. This representation organizes the interaction history of the terminal into explicit inputs for timely access decisions.

\subsection{Residual Policy Network}

The feature-level identity path in Fig.~\ref{fig:method}(a) connects the input and output of each residual block. The actor first maps the augmented input to a feature vector,
$\mathbf z^{(0)}_{i,t}
=
\phi(W_0\mathbf x_{i,t}+b_0),$
where $\mathbf z^{(0)}_{i,t}$ is the encoded input feature vector, $\phi(\cdot)$ is an elementwise activation function, and $W_0$ and $b_0$ are the trainable encoding weight matrix and bias vector, respectively.

Let $L_{\rm res}$ be the number of feature-residual blocks~\cite{resnet}. For block index $\ell=0,\ldots,L_{\rm res}-1$ and input feature vector $\mathbf z$, a two-layer branch computes
\begin{equation}
\label{eq:residual_branch}
F_{\theta_\ell}(\mathbf z)
=
W_{\ell,2}\phi(W_{\ell,1}\mathbf z+b_{\ell,1})
+b_{\ell,2}.
\end{equation}
where $W_{\ell,1}$ and $W_{\ell,2}$ are the branch weight matrices, $b_{\ell,1}$ and $b_{\ell,2}$ are their bias vectors, and $\theta_\ell$ collects these parameters. Writing $\mathbf z^{(\ell)}_{i,t}$ for the input of block $\ell$, the branch output is added through an identity connection:
\begin{equation}
\label{eq:residual}
\mathbf z^{(\ell+1)}_{i,t}
=
\mathbf z^{(\ell)}_{i,t}
+
F_{\theta_\ell}
\bigl(\mathbf z^{(\ell)}_{i,t}\bigr),
\end{equation}
for $\ell=0,\ldots,L_{\rm res}-1$. The input and output widths of each block agree.

The identity path carries the encoded current observations and historical feedback, while the nonlinear branch learns transformations of their combined representation. Under dynamic competition, the same access history can have different implications for current service needs. For example, a channel with favorable historical outcomes may incur excessive switching delay for an urgent packet. The residual blocks provide a feature representation in which access experience can be interpreted jointly with current sensing, queue conditions, and remaining packet lifetime.

A gated recurrent unit (GRU)~\cite{gru} then updates the local hidden-state vector $\mathbf s_{i,t}$ from its previous value $\mathbf s_{i,t-1}$ and the final residual feature vector:
$\mathbf s_{i,t}
=
\operatorname{GRU}_{\theta}
\left(
\mathbf z^{(L_{\rm res})}_{i,t},
\mathbf s_{i,t-1}
\right).$
Explicit feedback memory and recurrence have different roles: the former maintains predefined per-channel outcome statistics, while the latter encodes sequential context across current observations, service states, and those statistics. Recurrent state is retained across an ongoing operation and reset when the environment is reset.

A linear action head maps the updated recurrent state to one logit per action index:
$\boldsymbol\ell_{i,t}
=
W_a\mathbf s_{i,t}+b_a,$
where $W_a$ and $b_a$ are the trainable action-head weight matrix and bias vector. The logit vector $\boldsymbol\ell_{i,t}\in\mathbb R^{1+2M}$ contains the unnormalized action scores, with $\ell_{i,t,a}$ denoting the component for action $a$. Applying a masked softmax~\cite{mask} with the protocol mask from Section~\ref{sec:system} gives
\begin{equation}
\label{eq:policy}
\pi_\theta
\left(
a\mid\mathbf x_{i,t},\mathbf s_{i,t-1}
\right)
=
\frac{
\mu_{i,t}(a)\exp(\ell_{i,t,a})
}{
\sum_{a'\in\mathcal A_i}
\mu_{i,t}(a')\exp(\ell_{i,t,a'})
}.
\end{equation}
At least the WAIT/CONTINUE index is legal. When an operation is underway, CONTINUE has probability one. At a free decision point, the distribution selects among waiting, sensing, and eligible attempts on the available channels.

The policy combines complementary evidence when selecting an action. Repeated failures accompanied by favorable sensed energy indicate different experienced conditions from recent successes with stale sensing. Channel-specific feedback makes these distinctions available to the actor, and residual feature transformations and recurrence learn how they relate to timely delivery. The resulting recurrent representation produces action logits, which the masked softmax converts into a distribution over feasible actions.

\subsection{Centralized Training and Decentralized Execution}

{Fig.~\ref{fig:method}(c)} shows how joint training information produces value targets and shared actor updates.

\paragraph{Value estimation and temporal credit assignment}
During training, the critic receives the concatenated joint-input vector $\mathbf X_t=[\mathbf x_{1,t};\ldots;\mathbf x_{N,t}]$ and its previous hidden-state vector $\mathbf c_{t-1}$. Its value function $V_\psi(\mathbf X_t,\mathbf c_{t-1})$ estimates the expected discounted team return, where $\psi$ collects the critic parameters.
Its input is constructed entirely from the augmented local observations of the controlled links. The reward is the timely-confirmation count in~\eqref{eq:reward}, replicated across agents for cooperative training.

For a rollout collected with the old policy, write $V_t^{\rm old}$ for the value of the critic in the unnormalized reward scale. 
The bootstrap indicator $b_t\in\{0,1\}$ equals one when a continuation value is included and zero at a true terminal transition.
 The one-step temporal-difference residual is
$\delta_t
=
r_t+\gamma b_tV_{t+1}^{\rm old}-V_t^{\rm old}.$
This residual compares the observed reward plus the estimated continuation value with the current value estimate.

Generalized advantage estimation (GAE)~\cite{gae} accumulates these residuals backward along the rollout:
\begin{equation}
\label{eq:gae}
\widehat A_t
=
\delta_t
+
\gamma\lambda_{\rm GAE}\chi_t\widehat A_{t+1},
\end{equation}
where $\widehat A_t$ is the team advantage estimate, $\lambda_{\rm GAE}\in[0,1]$ controls the weighting of later residuals, and $\chi_t\in\{0,1\}$ equals one when the trace continues to the next recorded transition in the same realization and zero when the trace ends. Adding the estimated advantage to the old value gives the return target for critic training:
\begin{equation}
\label{eq:return_target}
\widehat R_t
=
\widehat A_t+V_t^{\rm old}.
\end{equation}

At a rollout boundary, the final observation supplies the bootstrap value and the finite sampled trace ends. At an artificial environment time limit, the pre-reset final observation supplies the bootstrap value, while the trace and recurrent states reset before the next realization. A true terminal transition uses $b_t=0$. These rules preserve continuation-value estimates while separating independently reset episodes~\cite{pardo}.

The same physical-slot discount is applied while a link is sensing, switching, or transmitting. Consequently, those stages influence the return even though their only legal action may be CONTINUE. For a singleton action set, the masked distribution has zero entropy and unit selected-action probability, and the direct actor-loss term at that slot has no policy dependence. Its reward and value information still contribute to temporal credit assignment for earlier decisions.

\paragraph{Actor and critic objectives}
Within the MAPPO training framework, the actor uses a clipped proximal policy optimization (PPO) objective with entropy regularization~\cite{ppo}, and the critic learns normalized return targets~\cite{mappo}.

For each stored local sequence, let $\rho_{i,t}(\theta)$ be the ratio of the new and stored old probabilities of the selected action, evaluated using the corresponding recurrent inputs and legal-action masks. Clipping this ratio to $[1-\epsilon,1+\epsilon]$ gives
$\bar\rho_{i,t}
=\operatorname{clip}(\rho_{i,t},1-\epsilon,1+\epsilon)$,
where $\epsilon$ is the policy clipping threshold.

Let $\widetilde A_t$ denote $\widehat A_t$ standardized using the mean and standard deviation of the rollout advantages. The actor minimizes
\begin{equation}
\label{eq:ppo}
\begin{split}
\mathcal L_{\rm actor}
=
-\E_{i,t}\Bigl[
&\min
\left\{
\rho_{i,t}\widetilde A_t,\,
\bar\rho_{i,t}\widetilde A_t
\right\}+\beta\mathcal H(\pi_\theta)
\Bigr],
\end{split}
\end{equation}
where $\E_{i,t}$ denotes the empirical average over sampled agent--time pairs. The minimum uses the smaller of the original and clipped surrogate terms. The categorical entropy $\mathcal H(\pi_\theta)$ is computed over legal actions and encourages exploration with nonnegative weight $\beta$.

The critic uses a normalized, clipped Huber loss. Let $\widetilde R_t$ denote the return target $\widehat R_t$ normalized using the running mean and standard deviation of return targets. The value predictions and their training targets use the same normalization statistics.
Let $v_t$ be the current normalized value prediction and $v_t^{\rm old}$ its stored old counterpart. The clipped prediction limits the change from the old prediction using threshold $\epsilon_v$:
\begin{equation}
\label{eq:value_prediction_clip}
\bar v_t
=
v_t^{\rm old}
+
\operatorname{clip}
\left(
v_t-v_t^{\rm old},
-\epsilon_v,\epsilon_v
\right).
\end{equation}

For prediction error $e$, the Huber penalty with threshold $\kappa>0$ is
\begin{equation}
\label{eq:huber_loss}
\ell_\kappa(e)
=
\begin{cases}
e^2/2, & |e|\leq\kappa,\\
\kappa(|e|-\kappa/2), & |e|>\kappa.
\end{cases}
\end{equation}
This penalty is quadratic for small errors and linear for large errors. The value objective uses the larger penalty obtained from the current and clipped predictions:
\begin{equation}
\label{eq:value_clip}
\mathcal L_{\rm value}
=
\E_t\Bigl[
\max\bigl\{
\ell_\kappa(\widetilde R_t-v_t),
\ell_\kappa(\widetilde R_t-\bar v_t)
\bigr\}
\Bigr].
\end{equation}
The value optimizer minimizes $c_v\mathcal L_{\rm value}$, where $c_v$ is the value-loss coefficient. Actor and critic use separate optimizers, and their PPO, value-normalization, and recurrent-training settings are held fixed across matched component comparisons.

\paragraph{Complete procedure and execution}
Feedback vectors are computed online and stored with observations, actions, masks, old probabilities, rewards, and recurrent states. Optimization reuses the recorded vectors so that each action remains paired with the information available when it was selected. Feedback memory evolves according to deterministic online update rules, while actor and critic parameters are optimized through their respective objectives.

Training samples legal actions, while evaluation selects the maximum-probability legal action, consistently with the learned baselines. The complete procedure is summarized in Algorithm~\ref{alg:far}.

\begin{algorithm}[t]
\caption{Feedback-Aware Residual Multi-Agent Reinforcement Learning}
\label{alg:far}
\begin{algorithmic}[1]
\Require Parallel environments, interaction budget, and PPO settings.
\Ensure Shared actor parameters $\theta$ for decentralized execution.
\State Initialize actor $\theta$, critic $\psi$, local feedback memories, and recurrent states.
\While{the interaction budget is not exhausted}
    \For{each physical slot in the rollout}
        \State Form local inputs $\mathbf x_{i,t}$, legal-action masks, and joint critic input $\mathbf X_t$.
        \State Sample legal actions using~\eqref{eq:policy}.
        \State Record the selected channel and start time for each new attempt.
        \State Advance all links by one physical slot and obtain the shared reward and local feedback.
        \State Decay recent statistics and incorporate each newly completed attempt exactly once.
        \State Store inputs, actions, masks, old probabilities, rewards, and recurrent states.
        \If{an environment reaches its time limit}
            \State Bootstrap from its pre-reset final observation.
            \State Reset its feedback memory and recurrent states.
        \EndIf
    \EndFor
    \State Compute advantages using~\eqref{eq:gae} and return targets using~\eqref{eq:return_target}.
    \State Update the actor and critic with recurrent mini-batches using~\eqref{eq:ppo} and~\eqref{eq:value_clip}.
\EndWhile
\Statex \textbf{Execution:} Each link uses the shared actor, maintains its own feedback memory and recurrent state, selects the highest-probability legal action, and continues feedback updates.
\end{algorithmic}
\end{algorithm}

This procedure connects the two design components to the communication task. Local memory supplies explicit evidence of experienced service conditions, and the residual recurrent actor learns how to use that evidence under current traffic and protocol constraints. Centralized training supports cooperative policy learning through a shared timely-delivery objective. After deployment, each link executes its policy using its own observations and interaction history. Section~\ref{sec:experiments} evaluates the combined design and its component variants.

\section{Performance Evaluation}
\label{sec:experiments}

This section evaluates access performance under different network configurations, traffic patterns, and packet deadlines. We first compare FAR-MAPPO with five baseline methods, then examine its components and delivery performance across deadline classes.

\subsection{Experimental Setup}

\subsubsection{Network and Traffic Settings}

The reference configuration consists of $N=12$ controlled links,
$K=12$ external competing links, and $M=4$ orthogonal channels.
The mean packet arrival rate is $\lambda=40$ packets/s per controlled link. Transmitters and receivers are deployed in a
$100\times100$~m area, with an intended link distance between
10 and 25~m.

Channel gains follow~\eqref{eq:gain}. The path-loss exponent is three, and the reference gain is determined by free-space propagation at 2.4~GHz over a distance of 1~m. Shadowing has a standard deviation of 4~dB, and exponential power fading is refreshed every 10~ms. The transmit power is 0~dBm, the noise floor is $-107$~dBm, and the sensing-error standard deviation is 2~dB. Reception and channel assessment follow the models in Section~\ref{sec:system}.

Unless otherwise specified, controlled links generate bursty traffic with mean ON and OFF durations of 20 and 60~ms, respectively. The stationary ON probability is $1/4$. Consequently, the packet arrival probability during an ON slot is $4\lambda\Delta$, maintaining an overall mean arrival rate of $\lambda$. Each arriving packet independently receives a deadline of 20, 40, or 80~ms with equal probability.

External links have mean ON and OFF durations of 20~ms each. Their backoff windows and data-transmission durations are selected from $\{4,8,16\}$ and $\{2,4,8\}$ slots, respectively, and remain fixed within each realization. Table~\ref{tab:settings} summarizes the key parameters.


\begin{table}[t]
\centering
\footnotesize
\caption{Key system parameters.}
\label{tab:settings}
\setlength{\tabcolsep}{3pt}
\renewcommand{\arraystretch}{1.08}
\begin{tabular}{@{}p{.40\linewidth}p{.55\linewidth}@{}}
\toprule
Parameter & Setting \\
\midrule
Controlled links $N$
& 12; sweep: 8, 10, 12, 14, 16, 18, 20 \\

External links $K$
& 12; sweep: 4, 6, 8, 10, 12, 14, 16 \\

Orthogonal channels $M$
& 4; sweep: 2, 3, 4, 5, 6, 7, 8 \\

Per-link arrival rate $\lambda$
& 40 packets/s; sweep: 20, 30, 40, 50, 60, 70, 80 \\

Deployment area; link distance
& $100\times100$~m; 10--25~m \\

Transmit power; noise floor
& 0~dBm; $-107$~dBm \\

SINR / CCA thresholds
& 6~dB / $-90$~dBm \\

Physical slot $\Delta$
& 1~ms \\

Switch / sensing / CCA / data / ACK
& 1 / 1 / 1 / 4 / 1~ms \\

Queue capacity; service order
& 20 packets; earliest deadline first \\

Packet deadlines
& Equiprobable 20, 40, and 80~ms \\

Controlled ON/OFF means
& 20/60~ms \\

External ON/OFF means
& 20/20~ms \\
\bottomrule
\end{tabular}
\end{table}

\subsubsection{Compared Methods}

The main comparisons include the following five baselines.

\begin{itemize}
    \item \textbf{Random:}
Each link selects an attempt channel uniformly at random and applies a random cooldown after unsuccessful or busy feedback.

    \item \textbf{Fixed:}
Link $i$ is assigned the preferred channel
    $1+((i-1)\bmod M)$ and applies a random cooldown after
unsuccessful or busy feedback.

    \item \textbf{Greedy:}
An idle link with a nonempty queue selects the channel with the lowest local score and attempts transmission. The score combines the latest measured channel energy with a 3-dB switching penalty. An unmeasured channel uses a default energy of $-100$~dBm. Score clipping and the first-minimum tie-breaking rule are fixed across tests.

    \item \textbf{Independent proximal policy optimization (IPPO):}
IPPO~\cite{ippo} uses a recurrent local actor and a critic based on local observations. Actor parameters are shared across links, with link identity included in the input.

    \item \textbf{MAPPO:}
MAPPO~\cite{mappo} uses recurrent local actors and a centralized critic that receives the joint observations of the controlled links.
\end{itemize}


\subsubsection{Training and Evaluation Protocol}

The feedback-memory decay scales are $T_f=64$~ms and
$T_s=512$~ms. The normalization caps are $C_{\max}=64$,
$C^{(f)}_{\max}=16$, $A_{\max}=100$~ms, $L_{\max}=10$,
and $\tau_{\max}=20$~ms.

For each network configuration, the learning methods use
524,288 network-level physical time steps accumulated across
eight parallel environments. The rollout length is 256, the number of PPO epochs per update is 15, and the recurrent hidden dimension is 64. The discount factor and GAE parameter are
0.99 and 0.95, respectively. Five training seeds are used:
32001, 32043, 32089, 32137, and 32183. Policies are trained
separately for each configuration and evaluated using their final-budget checkpoints. The interaction budget is matched across methods within each configuration. Repeated training and held-out evaluation account for variation across learned policies and network realizations, an important consideration in reinforcement-learning comparisons~\cite{agarwal}.

Evaluation uses twenty held-out environment seeds,
730001--730020. All methods are evaluated using the same
environment-seed set at a given configuration. Learned policies select the highest-probability legal action, while randomized baselines retain their prescribed randomization.

\begin{figure}[t]
\centering
\includegraphics[width=\linewidth]{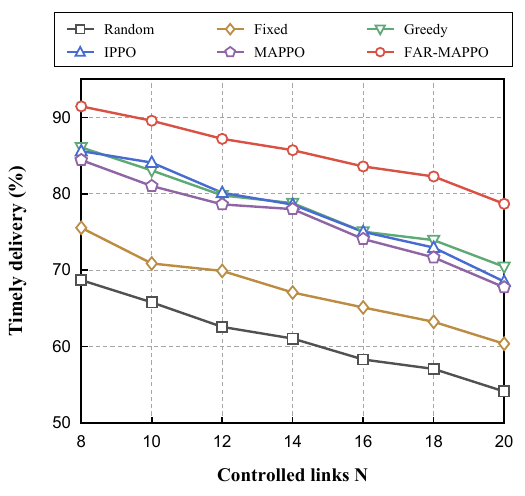}
\caption{
Timely delivery versus the number of controlled links, with $K=12$, $M=4$, and $\lambda=40$ packets/s/link.
}
\label{fig:n}
\end{figure}

Each evaluation realization includes a 2,000 ms warmup, a 10,000 ms measurement arrival window, and an 80 ms settling period. Arrivals and external competition continue during the settling period. Packet accounting verifies that every packet generated within the measurement window has been resolved. Packets discarded due to queue overflow or deadline expiry remain in the delivery-ratio denominator.

Let $S$ be the number of training seeds and $E$ the number of test environments. Using $s\in\{1,\ldots,S\}$ and $e\in\{1,\ldots,E\}$ as their enumeration indices, let $\eta_{s,e}$ denote the measured delivery ratio of the policy trained with seed $s$ in test environment $e$. The reported mean for a learning method is
\begin{equation}
\label{eq:eval_mean}
\widehat{\eta}
=
\frac{1}{S}
\sum_{s=1}^{S}
\left(
\frac{1}{E}
\sum_{e=1}^{E}
\eta_{s,e}
\right),
\end{equation}
where $S=5$ and $E=20$. Rule-based methods are averaged over the same twenty test environments. Numerical differences reported below are expressed in percentage points and are calculated before rounding.

Four parameter sweeps vary $N$, $K$, $M$, and $\lambda$ individually while keeping the other parameters at their reference values. Each sweep contains seven values, giving
25 distinct configurations after accounting for the shared
reference configuration. Increasing $N$ raises both aggregate traffic and the number of interacting links. Increasing $M$ provides additional channels while keeping per-channel propagation settings and packet-transmission duration unchanged.

\subsection{Performance under Different Network Configurations}

{Figs.~\ref{fig:n}--\ref{fig:load}} compare the methods across the four parameter sweeps. Over the 25 configurations, the mean timely-delivery ratio of FAR-MAPPO exceeds the strongest baseline mean by 2.59--9.38 percentage points.

\subsubsection{Number of Controlled Links}

{Fig.~\ref{fig:n}} varies $N$ from 8 to 20 with $K=12$,
$M=4$, and $\lambda=40$ {packets/s/link}. The aggregate arrival
rate therefore increases from 320 to 800 packets/s while the available channel resources remain fixed.

Greedy is the strongest baseline at both endpoints, with its delivery ratio decreasing from 86.06\% to 70.44\%. The corresponding FAR-MAPPO values are 91.44\% and 78.68\%, yielding improvements of 5.37 and 8.24 percentage points, respectively.

As the number of controlled links increases, more packets compete for transmission opportunities, and additional interference relationships emerge. The growing performance difference indicates the value of adapting access decisions to a larger interacting population under a fixed channel budget.

\subsubsection{Number of External Competing Links}

\begin{figure}[t]
\centering
\includegraphics[width=\linewidth]{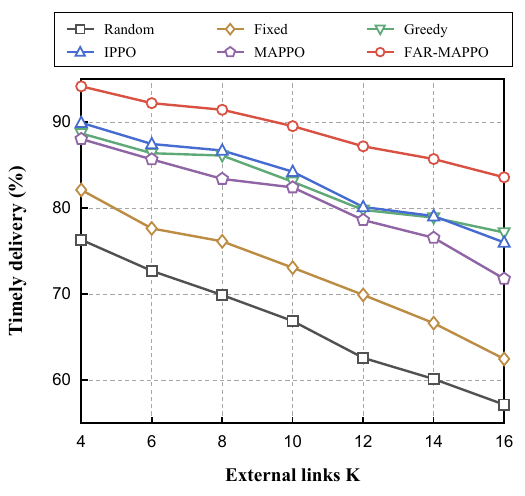}
\caption{
Timely delivery versus the number of external competing links, with $N=12$, $M=4$, and
$\lambda=40$ packets/s/link.
}
\label{fig:k}
\end{figure}

{Fig.~\ref{fig:k}} increases $K$ from 4 to 16 while fixing
$N=12$, $M=4$, and $\lambda=40$ packets/s/link.
At $K=4$, IPPO achieves the highest baseline delivery ratio of 89.93\%, compared with 94.16\% for FAR-MAPPO. At $K=16$, Greedy is the strongest baseline at 77.16\%, while FAR-MAPPO achieves
83.58\%. The corresponding improvements are 4.24 and
6.42 percentage points.

External competition affects both channel availability during CCA and reception quality during transmission. Consequently, local sensing and completed-attempt feedback provide complementary information for access decisions. The baseline ranking also changes with the competition level, showing that the relative effectiveness of sensing rules and learned policies depends on network conditions.

\subsubsection{Number of Orthogonal Channels}

\begin{figure}[t]
\centering
\includegraphics[width=\linewidth]{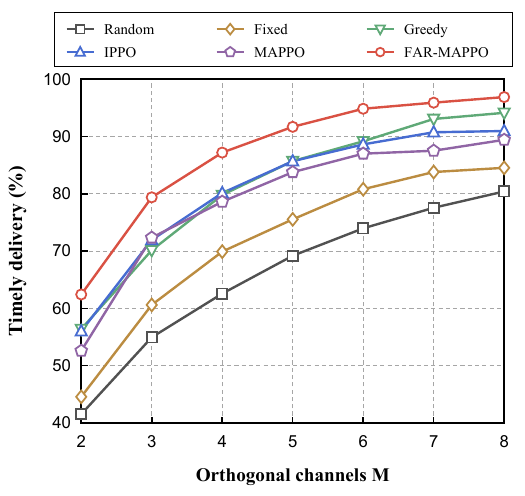}
\caption{
Timely delivery versus the number of orthogonal channels, with $N=K=12$ and $\lambda=40$ packets/s/link.
}
\label{fig:m}
\end{figure}

{Fig.~\ref{fig:m}} varies $M$ from 2 to 8 with $N=K=12$ and $\lambda=40$ packets/s/link. With two channels, Greedy achieves 56.39\%, whereas FAR-MAPPO achieves 62.42\%. With eight channels, their delivery ratios increase to 94.13\% and 96.87\%, respectively.

Additional channels allow concurrent transmissions to separate in frequency and reduce repeated contention for the same resource. The improvement over the strongest baseline is 6.03 percentage points at $M=2$, 7.06 points at $M=4$, and 2.75 points at $M=8$.

These results indicate that the performance benefit depends on the available service opportunities. Severe channel scarcity limits the delivery capability of all methods, while abundant channels allow the baselines to approach a high delivery ratio. The largest improvement among these three points occurs at an intermediate channel availability.

\subsubsection{Packet Arrival Rate}

\begin{figure}[t]
\centering
\includegraphics[width=\linewidth]{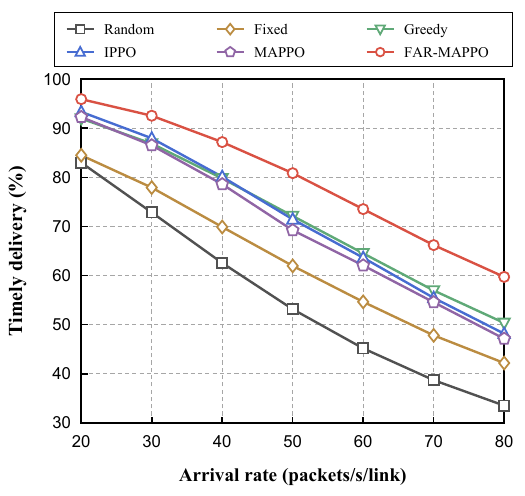}
\caption{
Timely delivery versus the mean per-link packet arrival rate, with $N=K=12$ and $M=4$.
}
\label{fig:load}
\end{figure}

{Fig.~\ref{fig:load}} varies the mean per-link arrival rate from 20 to 80 packets/s with $N=K=12$ and $M=4$. At 20 packets/s/link, IPPO achieves 93.34\%, compared with
95.93\% for FAR-MAPPO, giving a difference
of 2.59 percentage points. At 80 packets/s/link, Greedy provides the strongest baseline result of 50.36\%, while FAR-MAPPO achieves 59.73\%, increasing the difference to 9.38 percentage points.

Higher traffic loads place more packets within a limited service-time budget. Queueing, unsuccessful attempts, and protocol overhead consequently consume a larger share of the time available before packet expiry. The larger gain at high load shows that the performance difference becomes more pronounced when queueing and access overhead consume a greater share of packet lifetimes.

\subsection{Impact of Traffic Burstiness and Deadline Composition}

Table~\ref{tab:traffic} compares four traffic and deadline combinations at $N=K=12$, $M=4$, and
$\lambda=40$ packets/s/link. The arrival process is either
non-bursty or bursty, and packet deadlines are either uniformly 80~ms or independently selected from
$\{20,40,80\}$~ms with equal probability.

The mean arrival rate remains unchanged across these combinations. Burstiness therefore changes the temporal concentration of demand. Introducing mixed deadlines adds shorter packet lifetimes and reduces the mean deadline from 80 to 46.67~ms, jointly changing deadline composition and tightness.

\begin{table*}[t]
\centering
\footnotesize
\caption{
Timely delivery (\%) under different traffic and deadline combinations, with $N=K=12$, $M=4$, and
$\lambda=40$ packets/s/link.
}
\label{tab:traffic}
\setlength{\tabcolsep}{4pt}
\renewcommand{\arraystretch}{1.08}
\begin{tabular*}{.98\textwidth}
{@{\extracolsep{\fill}}lrrrrrr}
\toprule
Traffic and deadline
& Random
& Fixed
& Greedy
& IPPO
& MAPPO
& FAR-MAPPO \\
\midrule
Non-bursty, 80~ms
& 89.16 & 88.59 & 93.51 & 96.59 & 92.63 & 98.54 \\

Non-bursty, mixed
& 77.87 & 80.85 & 88.62 & 88.93 & 87.17 & 93.59 \\

Bursty, 80~ms
& 74.33 & 79.33 & 87.56 & 88.95 & 85.30 & 93.74 \\

Bursty, mixed
& 62.57 & 69.90 & 79.78 & 80.13 & 78.61 & 87.19 \\
\bottomrule
\end{tabular*}
\end{table*}

IPPO is the strongest baseline in all four combinations. Its delivery ratio is 96.59\% with non-bursty arrivals and an 80-ms deadline. Introducing mixed deadlines reduces this value to 88.93\%, while introducing bursty arrivals with the 80-ms deadline yields 88.95\%. Combining burstiness and mixed deadlines further reduces the delivery ratio to 80.13\%.

The corresponding FAR-MAPPO results are
98.54\%, 93.59\%, 93.74\%, and 87.19\%.
Its improvements over the strongest baseline are
1.95, 4.67, 4.78, and 7.06 percentage points,
respectively.

The same average offered load can thus impose substantially different timely-service requirements. Concentrated arrivals increase short-term competition for service, while shorter deadlines reduce the time available to recover from waiting or unsuccessful attempts. Their combination produces a larger separation between the compared access strategies.

\subsection{Ablation Study}

Table~\ref{tab:ablation} compares MAPPO, Feedback-Aware MAPPO
(FA-MAPPO), Residual-Network MAPPO (ResNet-MAPPO), and
FAR-MAPPO. FA-MAPPO augments the ordinary MAPPO architecture with historical feedback, ResNet-MAPPO introduces the residual policy network, and FAR-MAPPO combines both components.

All three configurations use $N=K=12$, with
$(M,\lambda)$ set to $(4,40)$, $(2,40)$, and $(4,80)$,
where $\lambda$ is measured in packets/s/link. They represent the reference configuration, reduced channel availability, and increased traffic load, respectively.

\begin{table}[t]
\centering
\footnotesize
\caption{
Component-comparison means: timely delivery (\%) with
$N=K=12$.
}
\label{tab:ablation}
\setlength{\tabcolsep}{12pt}
\renewcommand{\arraystretch}{1.12}
\begin{tabular}{@{}lrrr@{}}
\toprule
Method
& \shortstack{$M=4$\\$\lambda=40$}
& \shortstack{$M=2$\\$\lambda=40$}
& \shortstack{$M=4$\\$\lambda=80$} \\
\midrule
MAPPO
& 78.61 & 52.59 & 47.14 \\

FA-MAPPO
& 81.07 & 60.49 & 57.54 \\

ResNet-MAPPO
& 86.72 & 62.18 & 58.28 \\

FAR-MAPPO
& 87.19 & 62.42 & 59.73 \\
\bottomrule
\end{tabular}
\end{table}

Relative to MAPPO, FA-MAPPO improves the mean timely-delivery ratio by 2.47, 7.90, and 10.40 percentage points across the three configurations. ResNet-MAPPO improves the mean timely-delivery ratio by
8.11, 9.59, and 11.14 points.
FAR-MAPPO achieves the highest mean timely-delivery ratio in each configuration, exceeding MAPPO by
8.58, 9.83, and 12.59 points, respectively.

The comparison with each individual-component row further characterizes the combined result. FAR-MAPPO exceeds FA-MAPPO by
6.11, 1.93, and 2.19 percentage points, and exceeds
ResNet-MAPPO by 0.47, 0.24, and 1.46 points. Thus, the difference from the original MAPPO backbone is larger than the additional improvement over the stronger individual-component row.

The method design assigns complementary roles to the two components: historical feedback provides channel-specific evidence about completed access attempts, while residual feature processing learns how this evidence and the current local state should influence the decision.

\subsection{Delivery Performance across Deadline Classes}

\begin{figure*}[t]
\centering
\includegraphics[width=\textwidth]{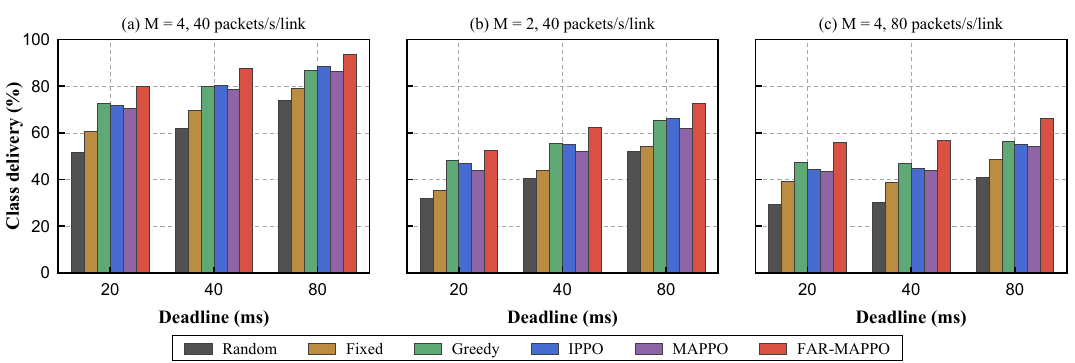}
\caption{
Timely delivery across deadline classes with $N=K=12$:
(a) $M=4$ and $\lambda=40$ packets/s/link;
(b) $M=2$ and $\lambda=40$ packets/s/link;
(c) $M=4$ and $\lambda=80$ packets/s/link.
Each delivery ratio uses all generated packets in the corresponding deadline class as its denominator.
}
\label{fig:deadlines}
\end{figure*}

{Fig.~\ref{fig:deadlines}} reports timely delivery separately for packets with {20 ms, 40 ms, and 80 ms} deadlines under the same three configurations as Table~\ref{tab:ablation}. For each deadline class, the denominator includes all packets generated in that class. This comparison examines how the aggregate delivery pattern extends to packets with different time budgets.

At $M=4$ and $\lambda=40$ packets/s/link, FAR-MAPPO achieves 80.12\%, 87.79\%, and 93.70\% for the 20-, 40-, and 80-ms classes, respectively. For the shortest-deadline class, Greedy is the strongest baseline at 72.69\%, giving an improvement of 7.43 percentage points.

When the number of channels decreases to two, the 20-ms delivery ratio of FAR-MAPPO is 52.42\%, compared with 48.17\% for the strongest baseline. At $\lambda=80$ packets/s/link with four channels, the corresponding values are 55.90\% and 47.38\%. These differences are 4.25 and
8.52 percentage points, respectively.


\section{Conclusion and Future Work}
\label{sec:conclusion}

This paper investigated timely distributed communication in cognitive radio networks with partially observable reception conditions, dynamic competition, and deadline-consuming access procedures. We developed FAR-MAPPO, which combines historical feedback representation with a residual policy network under centralized training and decentralized execution. This design integrates channel-specific access experience with current service conditions to guide local decisions. The numerical results show improved mean timely delivery across the evaluated configurations, with larger gains under heavy traffic and bursty arrivals with mixed deadlines. The improvements also extend to short-deadline packets.

Future work will extend this framework toward joint channel selection, transmission scheduling, and power control, enabling coordinated use of spectrum, time, and energy resources. We will also investigate transferable feedback representations that facilitate policy reuse across network scales and service profiles. Building on the current delivery objective, another direction is to incorporate differentiated service priorities and energy budgets into multi-agent coordination. These extensions would support a broader range of wireless applications while retaining the use of local feedback by the framework for decentralized decision-making.

\bibliographystyle{IEEEtran}
\bibliography{references}
\end{document}